\documentclass[useAMS,usenatbib]{mn2e}
\usepackage{graphicx}    
\usepackage{dcolumn}     
\usepackage{bm}          
\usepackage{amssymb,amsmath}  
\usepackage{color}
\usepackage{fixltx2e}
\usepackage{lastpage}
\usepackage{xfrac, multirow, url, balance, placeins}
\usepackage{amsfonts}
\usepackage{rotating}
\usepackage{widetext}
\usepackage[none]{hyphenat}
\usepackage[english]{babel}

\usepackage[stable]{footmisc}

\title[Nonparametric reconstruction of the $O_m$ diagnostic to test $\Lambda$CDM]{Nonparametric reconstruction of the $O_m$ diagnostic to test $\Lambda$CDM}  

\author[Celia Escamilla-Rivera and Julio Fabris]{Celia Escamilla-Rivera$^{1}$\footnotemark[0]\thanks{Email: cescamilla@mctp.mx},  
Julio Fabris$^{2,3}$\footnotemark[0]\thanks{Email: fabris@pq.cnpq.br}
\\\\\\
$^{1}$Mesoamerican Centre for Theoretical Physics (MCTP), Universidad Aut\'onoma de Chiapas, 29040, Tuxtla Guti\'errez, Chiapas, M\'exico. \\
$^{2}$Departamento de Fisica, Universidade Federal do Espirito Santo,
Av. Fernando Ferrari 514, Vitoria, ES, 29075-910, Brazil.\\
$^{3}$National Research Nuclear University "MEPhI", Kashirskoe sh. 31, Moscow 115409, Russia.
}

\begin{document}

\date{\today}
\pagerange{\pageref{firstpage}--\pageref{LastPage}} \pubyear{2016}
\maketitle

\label{firstpage}

\begin{abstract}
Cosmic acceleration is usually related with the unknown dark energy, which
equation of state, $w(z)$, is constrained and numerically confronted with independent 
astrophysical data. 
In order to make a diagnostic of $w(z)$,
the introduction of a null test of dark energy can be done using a diagnostic function of redshift, $O_m$.
In this work we present a nonparametric reconstruction of this diagnostic using the so-called 
Loess-Simex factory to test the concordance model with the advantage 
that this approach 
offers an alternative way to relax the use of priors and find a possible $w$ that reliably describe the
data with no previous knowledge of a cosmological model. Our results 
demonstrate that the method applied to the dynamical $O_m$ diagnostic finds a preference for 
a dark energy model with equation of state $w=-2/3$, which correspond to a static domain wall network.
\end{abstract}



\section{Introduction}

At present numerous projects and surveys are either underway or being proposed [\cite{surveys}] to
discover the underlying cause of the accelerated expansion which is well established by present 
observations as: Supernovae Type Ia (SNIa) [\cite{Riess:1998cb-Perlmutter:1998np}], Baryon 
Acoustic Oscillations (BAO) [\cite{Eisenstein:2005su}],  Cosmic Microwave Background 
Radiation (CMBR) anisotropies [\cite{Spergel:2003cb}], Large Scale Structure formation [\cite{Tegmark:2003ud}] 
and Weak Lensing [\cite{Jain:2003tba}]. The current standard cosmological model,
consistent with these vast observations, is the $\Lambda$CDM or concordance model, 
in which this accelerated behaviour is driven by a cosmological constant $\Lambda$ and filled 
with Cold Dark Matter (CDM). This $\Lambda$ is usually related to an extra component
in the Universe, the so-called Dark Energy (DE) with $w=-1$. 
Despite of its simplicity, the $\Lambda$CDM
model has a couple of theoretical loopholes (e.g the fine tuning and coincidence problems
[\cite{Perivolaropoulos:2011hp}] which had lead to alternative proposals that 
either modified the General 
Relativity or consider a landscape with a dynamic DE. In this way, DE can
be described by an equation of state (EoS) written in terms of the redshift, $w(z)$, but until now, we 
do not have a precise evidence and/or evolution of this quantity. Since its properties are still under research, 
a wide zoo of reconstructions of DE parameterizations has been proposed to help to discern on the 
dynamics of this component [\cite{demodels}].

Despite the efforts to solve the theoretical loopholes of the concordance model there has been no strong alternative yet. 
In this matter, it results useful to test the consistency $\Lambda$CDM with cosmological observations 
and comparing it with the alternatives models or parameterizations. However, this mainstream is 
unlike to give any new physics beyond this scenario, 
but to reveal such possible new physics is essential to avoid
a prior knowledge of a cosmological model in order to find an adequate EoS that reliable describe the
astrophysical data available.
An important goal in the same line is to differentiate $\Lambda$CDM model from others
DE models in a scenario that has the less priors as possible, because as we have experienced over the 
years, incorrect priors of $w(z)$ or values of the density quantities 
can lead us to incorrect cosmological results. 
An interesting null test of DE, called $O_m$ diagnostic, was proposed in [\cite{Sahni:2008xx}]. The
elegance of this proposal lies in its theoretical form, which is constructed using only the Hubble parameter
$H(z)$, quantity that can be measure directly from the observations. This
procedure allows differentiate between the cosmological constant (flat $\Lambda$CDM) from a 
dynamical model (curved $\Lambda$CDM) only by considering as a prior the value of $\Omega_{m}$. 
Even if the value of $\Omega_m$ is not accurately known, the authors of the previous reference
gives some interesting insights in [\cite{Shafieloo:2012rs}] using an extension of the $O_m$ diagnostic called
\textit{two-points difference}. As a step forward, 
in [\cite{Seikel:2012cs}] was analyzed a curved $\Lambda$CDM, in where the diagnostic function 
$O^{(2)}_m$ includes first derivatives of $H(z)$ and a new parameter related to the curvature, $O_k$, enters 
to the scene. These tests 
are quite helpful 
because we have a scenario in where the diagnostic function can tell us if the previous DE assumptions are in agreement 
with the $\Lambda$CDM model or deviates from it towards an alternative DE or a modified gravity model.

One of the most useful astrophysical 
tool used is the luminosity distance of SNIa observations, which had the advantage to lead to 
$H(z)$ via the first derivative of this quantity.
So far, there are two astrophysical samples that reflect directly measures of it:
first, the Cosmic Chronometers (C-C), which gives a compilation
of $H(z)$ measurements estimated with the differential evolution of passively evolving early-type galaxies 
[\cite{Hsamples}];
second, the radial BAO scale in the galaxy distribution, a relic of the pre-recombination universe 
[\cite{Blake:2012pj,Gaztanaga:2008xz}]. The mentioned diagnostic has been tested with these 
astrophysical samples and provide a solution of the cosmic acceleration based in a smoothed model-independent  
via Gaussian process [\cite{Holsclaw:2010sk,Shafieloo:2012ht}], but the price that we pay for using this are
the strong constraint over the statistical process 
and the assumption of a initial guess cosmological model.

In the light of these issues, in [\cite{Montiel2013}] was proposed the use of two statistical techniques: 
the Locally Weighted Scatterplot Smoothing (Loess) [\cite{cleveland_loess}] and the Simulation and 
Extrapolation methods (Simex) [\cite{carroll_simex}]
in order to address a nonparametric scenario with the fewest possible of priors,
a smooth reconstruction of the parameter $H(z)$ and, of course, obtain the well established cosmic 
acceleration. Two nobel achievements using these statistical techniques are: \textbf{(1)} \textit{we 
do not need any DE parameterization as a prior}, instead we apply directly the full astrophysical
sample in the code structure and the evolution of the cosmological parameters will be issued 
by the smooth curve given by the observations;
\textbf{(2)} \textit{we do not require any functional distribution for the analysis}. There is only a couple of
restrictions which are related to the statistical analysis: \textbf{(a)} the size of the window data in where we are going to 
develop a fitting routine based in a specific degree of the polynomial [\cite{numerical,Daly:2003iy}]; 
\textbf{(b)} and we require a weight function that will gives to each data point some importance with respect to the other observations 
around them.
We clarify that this factory is a cosmological-model-independent method due the relax in the use of 
information concerning cosmological parameters in comparison to Gaussian process, where the use
of strong constraints on spatial flatness is required [\cite{Holsclaw:2010sk}].
In order to proceed with this research, we will follow these ideas to constraint even more
the use of priors via the Loess-Simex factory and reconstruct $h(z)$ and its derivative to test the 
$\Lambda$CDM model. 

This paper is organised as follows.
In Sect. \ref{sec:backgroundeqs} we give an overview of 
the quantities used to test the $\Lambda$CDM model. 
In Sect. \ref{sec:omintro} and \ref{sec:omdynamical} we
derive the equations for the $O_m$ diagnostic by consider a constant EoS and present the cases 
for a flat and curved universe. 
In Sect. \ref{sec:observations} we describe the astrophysical samples for $H(z)$.
In the following two sections we describe our methodology with the Loess-Simex factory to reconstruct 
$h(z)$ and the $O_m$ diagnostic.
We conclude in Sect. \ref{sec:interpretation} with a discussion of the results obtained.


\section{$\Lambda$CDM background}
\label{sec:backgroundeqs}

The dark energy reconstruction starts underlying the validity of the FLRW metric which gives the 
Friedmann equation
\begin{eqnarray}\label{eq:friedmann}
\left(\frac{H(z)}{H_{0}}\right)^2 \equiv h^2 (z)&=&\Omega_{0m} (1+z)^3 +\Omega_{0k} (1+z)^2 \nonumber\\&&
+(1-\Omega_{0m} -\Omega_{0k})f(z),
\end{eqnarray}
where
\begin{equation}\label{eq:fz}
f(z)=\exp\left[3\int^{z}_{0}{d\tilde{z} \left(\frac{1+w(\tilde{z})}{1+\tilde{z}}\right)}\right],
\end{equation}
and $\Omega_{0m}$, $\Omega_{0k}$ are the matter and curvature densities at present epoch, respectively.
The EoS that characterize DE can be obtained by introducing Eq.(\ref{eq:fz}) in
Eq.(\ref{eq:friedmann}) and deriving to obtain its characteristic expression
\begin{equation}\label{eq:de_eos}
w(z)=\frac{2(1+z)hh'-3h^2 +\Omega_{0k} (1+z)^2}{3\left[h^2 -\Omega_{0m} (1+z)^3 -\Omega_{0k} (1+z)^2\right]},
\end{equation}
where $h'(z)$ is the first derivative of the normalized Hubble parameter with respect to the redshift $z$.
Here we can notice that depending the values of the density parameters there is a strong restriction over $w(z)$.
The simplest explanation for DE is when this parameter acquire
the value $w=-1$, which is related to a cosmological constant $\Lambda$. Another interesting cases emerge
when $w>-1$ ($w<-1$), which points out to quintessence (phantom) scenario, respectively. However, the models 
are still restricted to the values of the density parameters and a distinction between them are quite difficult at this point.

This issue was the pattern to propose a diagnostic to differentiate between DE models in scenarios where $w$ could be a 
constant (and flat) and dynamical (and non-flat). 
$O_m$ diagnostic outline a test where we can fathom between DE models in the cases when the
value of $O_m$ is a constant or not. Following these lines, let us start our study by describing a $O_{m}$ diagnostic 
with a flat $\Lambda$CDM model as an example. Afterwards, we will proceed with the presentation of the
 dynamical (non-flat) diagnostic. 

\section{The $O_m$ diagnostic background}
\label{sec:omintro}

Let us begin with the distance-redshift relation
\begin{equation}\label{eq:dist-red1}
D(z)=\frac{H_0}{c}\frac{d_L (z)}{(1+z)},
\end{equation}
where 
\begin{equation}\label{eq:lum_dist1}
d_L (z)=\frac{c(1+z)}{H_0 \sqrt{-\Omega_{0k}}}\sin{\left[\sqrt{-\Omega_{0k}}\int^{z}_{0}d\tilde{z} \frac{H_0}{H(\tilde{z})}\right]},
\end{equation}
is the luminosity distance. Deriving Eqs.(\ref{eq:dist-red1})-(\ref{eq:lum_dist1}) and consider
a flat universe ($\Omega_{0k} =0$), it can be found that $D'(z)=H_0/H \equiv h^{-1}$. In this flat background with 
a constant DE EoS, $w=w_0$, the Eq.(\ref{eq:friedmann}) can be expressed as:
\begin{equation}\label{eq:hdef}
h^2 (z)=\Omega_{0m}(1+z)^3 +(1-\Omega_{0m})(1+z)^{3(1+w_0)},
\end{equation}
from where we can define a function that characterize this diagnostic 
\begin{equation}\label{eq:omgen}
O^{(1)}_{m_{0}}(z)\equiv\frac{h^2 - (1+z)^{3(1+w_0)}}{(1+z)^3 \left[1-(1+z)^{3w_0}\right]},
\end{equation}
where the upper index `(1)' indicates the existence of a first derivative of the luminosity distance $d_L$. 

\begin{table}
\begin{center}
\begin{tabular}{|c |c |c |}
  \multicolumn{2}{c}{}\\
  \hline
  EoS&  $O_m$ diagnostic  &  Model \\
  \hline
  $w_0=-1$ & $O^{(1)}_{m} =\Omega_{0m}$ & Flat $\Lambda$CDM.\\
 $w_0>-1$ &  $O^{(1)}_{m} >\Omega_{0m}$ & Quintessence. \\
 $w_0<-1$ &  $O^{(1)}_{m} <\Omega_{0m}$ &  Phantom.\\
  \hline
\end{tabular}
\end{center}
  \caption[x]{Features in the $O_m$ diagnostic with respect to the value of $\Omega_{0m}$, which can be taken 
  from recent Planck results [\cite{Ade:2015xua}] and a constant EoS $w=w_0$.}
  \label{tablecosmo}
\end{table}

\begin{figure}
\vspace{-1em}
\includegraphics[width=0.43\textwidth]{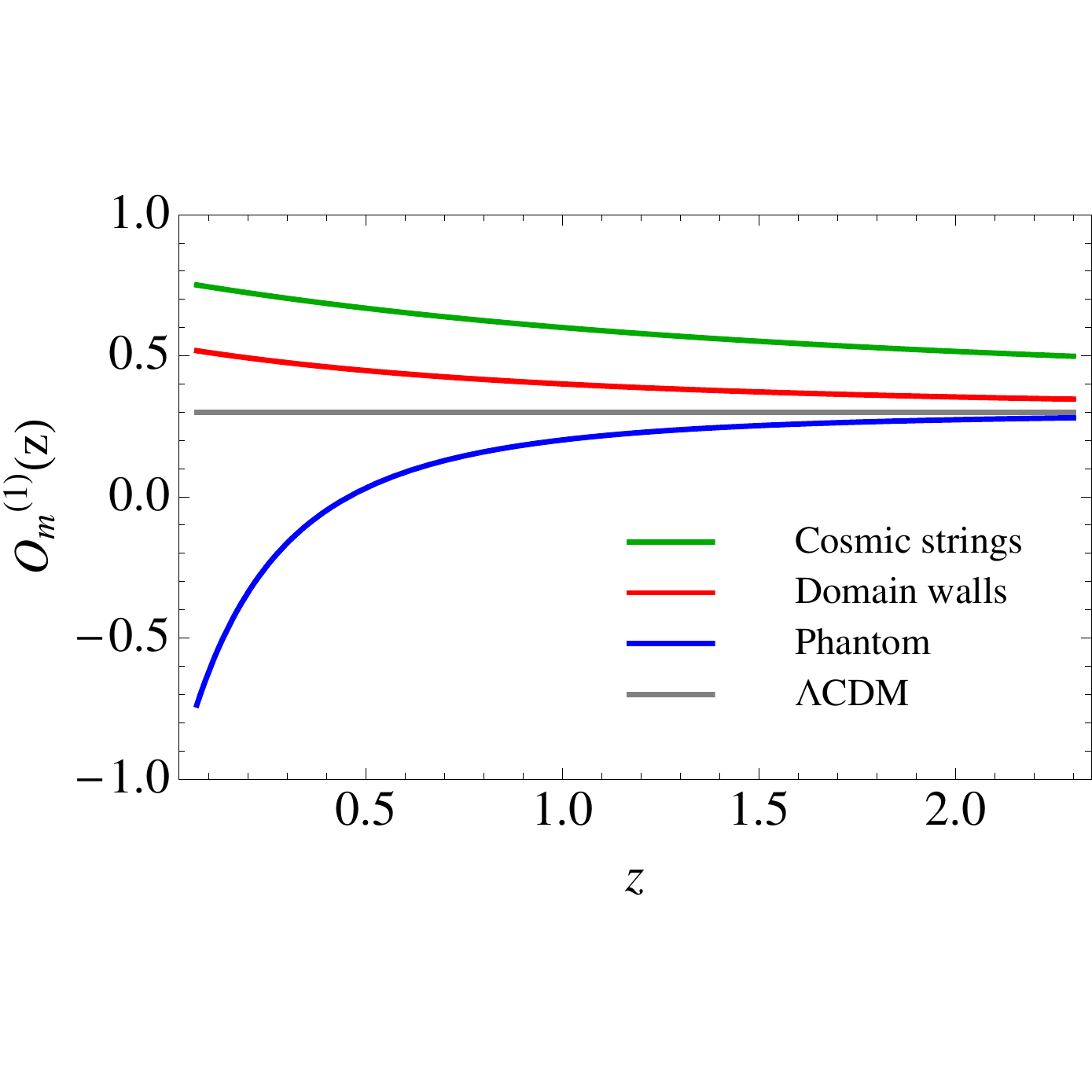}
\caption{Comparison between dark energy models. The solid grey line represent a $\Lambda$CDM model with $\Omega_m =0.315$ 
}
\label{fig:plots_comparison}
\end{figure}

To test the $\Lambda$CDM model using direct observations of the Hubble rate $H(z)$,
we require set in Eq.(\ref{eq:omgen}) $w_0 =-1$ [\cite{Sahni:2008xx}]
\begin{equation}\label{eq:omega1}
O^{(1)}_{m}(z)=\frac{h^2 -1}{z(3 +3z +z^2)}.
\end{equation} 

At this point, we can distinguish a $\Lambda$CDM model from any DE models by rewriting Eq.(\ref{eq:omega1}) using
Eq.(\ref{eq:hdef}), obtaining  
\begin{equation}\label{eq:finalom}
O^{(1)}_{m}(z)= \Omega_{0m} +(1-\Omega_{0m})\left[\frac{(1+z)^{3(1+w_0)}-1}{(1+z)^3 -1}\right],
\end{equation}
in where, on one hand, if $w_0 =-1$ implies $\Lambda$CDM with $O^{(1)}_{m}= \Omega_{0m}$. On the
other hand, if $w_0 > -1$ (or $w_0 < -1$) implies quintessence (or phantom) scenarios with $O^{(1)}_{m}> \Omega_{0m}$
 (or $O^{(1)}_{m}< \Omega_{0m}$), respectively. These descriptions are detailed in Table \ref{tablecosmo}.
 
Once we consider a Hubble rate $h(z)$ sample is possible to estimate confidence values of $O^{(1)}_{m}$. If
the test does not give a constant behaviour 
then the $\Lambda$CDM model is rule out and the existence of DE models or a curved $\Lambda$CDM scenario are considered.
In the first option, several DE candidates can be related to $O^{(1)}_{m}$ (see Figure \ref{fig:plots_comparison}) by considering
a specific value for $w_0$, e.g non-interacting cosmic string with $w_0 =-1/3$ [\cite{Alam:2003sc}] and static domain walls with $w=-2/3$ [\cite{Friedland:2002qs}]. 
To distinguished between these models we require the introduction of the $O_m$ diagnostic at first-order in $h$, which is related to the
dynamical test.
 

\section{The dynamical $O_m$ diagnostic}
\label{sec:omdynamical}

A more meticulous analysis based in the above-mentioned features takes into account a curved model,
where first 
derivatives of $h(z)$ comes into the scene. Expressions for this case can be obtained by consider
$\Omega_{0k} \neq 0$ and $w=w_0$ in Eq.(\ref{eq:friedmann}):
\begin{equation}\label{eq:hdef2}
h^2 (z) =\Omega_{0m} (1+z)^3 +\Omega_{0k} (1+z)^2 +(1-\Omega_{0m} -\Omega_{0k})(1+z)^{3(1+w_0)},
\end{equation}
from where we can find two expressions:
\begin{widetext}
\begin{eqnarray}\label{eq:omgen2}
O^{(2)}_{m_0} (z) &=& \frac{h^2 \left[3w_0 (1+z)^{3w_0+1} +3z(1+z)^{3w_0} +3(1+z)^{3w_0}-2\right] -(1+3w_0)(1+z)^{3(1+w_0)} }{(1+z)^3 \left[1-(1+z)^{3w_0} +3w_0 z(1+z)^{3w_0}\right]} \nonumber\\&&
-\frac{2h h' (1+z) \left[z(1+z)^{3w_0}+(1+z)^{3w_0}-1\right]}{(1+z)^3 \left[1-(1+z)^{3w_0} +3w_0 z(1+z)^{3w_0}\right]}, 
\end{eqnarray}
\begin{eqnarray}
O_{k_0} (z) = \frac{3\left\{w_0 (1+z)^{3(1+w_0)}-h^2 \left[w_0 (1+z)^{3w_0}+(1+z)^{3w_0} -1\right]\right\} +2h h' (1+z)\left[(1+z)^{3w_0 }-1\right]}{(1+z)^2 \left[3w_0 z (1+z)^{3w_0} -(1+z)^{3w_0} +1\right]}, 
\label{eq:omgen2_1}
\end{eqnarray}
\end{widetext}
where the upper index `(2)' indicates the existence of a second derivative of the luminosity 
distance. The calculations are explained in Appendix \ref{GE}.

We can obtain the curved $\Lambda$CDM case when we consider $w_0 =-1$ in the system (\ref{eq:omgen2})-(\ref{eq:omgen2_1}), which 
gives [\cite{Seikel:2012cs}] 
\begin{eqnarray}
O^{(2)}_{m}(z)&=&\frac{2\left[(1+z)(1-h^2) +z(2+z)hh'\right]}{z^2 (1+z)(3+z)}, \label{eq:omega2}\\
O_{k}(z)&=& \frac{3(1+z)^2 (h^2 -1)-2z(3+3z+z^2)hh'}{z^2 (1+z)(3+z)}.\label{eq:omegak} \quad\quad
\end{eqnarray}

To perform the distinctions between DE models we can rewrite Eq.(\ref{eq:omgen2}) using 
Eq.(\ref{eq:hdef2}) and its derivative, which
gives $O^{(2)}_{m_0}= \Omega_{0m}$ and $O_{k_0}= \Omega_{0k}$ implying a $\Lambda$CDM model.

\section{Observations of the Hubble rate}
\label{sec:observations}
To perform the diagnostic analysis we require to have at hand the observed
$H(z)$ data. This parameter has become an effective probe 
in cosmology 
comparison with SNIa, BAO and CMB data. In fact, it is more rewarding to study
the observational $H(z)$ data directly due
that all these tests use the distance scale
(e.g the luminosity distance $d_L$, the shift parameter $R$, or the distance parameter $A$)
measurement to determinate the values of the cosmological parameters, which needs the integral of $H(z)$
and therefore loses some important information of this quantity. 

$H(z)$ depends on the differential age as a function of redshift $z$ in the form: $H(z)=-(1+z)^{-1} dz/dt$, which
gives a direct measurement of $H(z)$ through the change of redshift in cosmic time. As an independent 
approach of this measure we provide two samples:
\begin{enumerate}
\item \textit{Cosmic Chronometers (C-C) data}. This kind of sample gives a measurement of the expansion 
rate without relying on the nature of the metric between the chronometer and us. We are going to employ  
several data sets presented in [\cite{Hsamples}]. A full compilation of the latter, which include 28 measurements 
of $H(z)$ in the range $0.07 < z < 2.3$, are reported in [\cite{Hsamples2}]. The normalized parameter $h(z)$ can be easily 
determine by consider the value $H_0 = 67.31 \pm 0.96$ kms${}^{-1}$ Mpc${}^{-1}$ [\cite{Ade:2015xua}]. 

\item \textit{Data from BAO}. Unlike the angular diameter $d_A$ measures given by the transverse BAO
scale, the $H(z)$ data can be extracted from the measurements of the line-of-sight of this BAO scale. 
Because the BAO distance scale is embodied in the CMB, its measurements on DE parameters strongest
at low redshift. The samples that we are going to consider consist in 3 data points from [\cite{Blake:2012pj}] 
and 3 more from [\cite{Gaztanaga:2008xz}] measured at 6 redshift in the range $0.24 < z < 0.73$. 
This data set is showed in Table \ref{tab:dataBAO}.
\begin{table}
\caption{\textit{\textit{BAO}} sample data from
}\label{tab:dataBAO}
\centering
\resizebox*{0.32\textwidth}{!}{
\begin{tabular}{ccccc}
\hline
\hline
{z}  &  {$H(z)$ [km$\text{s}^{-1}\text{Mpc}^{-1}$]}  & {${\sigma_{H}}^2$} \\
\hline
\hline
$0.24$  & $79.69$   & $2.32$  \\
$0.34$  & $83.80$   & $2.96$  \\
$0.43$  & $86.45$   & $3.27$ \\
$0.44$   & $82.6$   & $7.8$  \\
$0.6$  & $87.9$   & $6.1$  \\
$0.73$  & $97.3$   & $7.0$ \\
\hline
\hline
\end{tabular}}
\end{table}
\end{enumerate}

\section{Nonparametric reconstructions}
\label{sec:loess-simex-model}

Following the same methodology proposed in [\cite{Montiel2013}], we are going to 
reconstruct the normalized Hubble parameter $h$ using the Loess-Simex factory. 

\subsection{Reconstruction of $h(z)$}

\textbf{Step A1}. \textit{Windows and subsample selection.} First, we are going to select the 
proportion of observations fitting in a specific window. 
Each selection consist in some percentage of the total number of 
observations and to each subsample will be assigned a specific weighted least square local 
polynomial fit. We use a subsample via one quantity that is
usually known in the statistical 
jargon as the smoothing parameter or \textit{span} $s$, 
we use $k=ns$, where $k$ is the number of observations per window and rounded to
the next largest integer, $n$ is the 
total number of observations and $s$ typically takes values that oscillates between 0 and
1. We calculated the values: $s=0.9$ for the C-C sample, $s=0.85$ for the BAO
 sample and $s=0.4$ for the C-C+BAO total sample, which correspond to 90, 85 and 40 percent of the data in each window, respectively. 
These values were found using the \textit{cross validation} process detailed in [\cite{Montiel2013}].

\textbf{Step A2}. \textit{Weighted subsamples.} Having already selected the amount of data 
in each window, consider a certain amount of
data points near each other are more related between them than others that are significantly away
and receive a null weight. This idea is coined in the weight function described by a \textit{tricube kernel}: 
\begin{equation}\label{eq:tricube}
   W(\bar{z}_{i}) = \left\{
     \begin{array}{lr}
       \left(1-|\bar{z}_{i}|^3\right)^3 & \text{for} \quad |\bar{z}_{i}|<1,\\
       &\\
       0 &    \text{for}  \quad |\bar{z}_{i}|\ge1,
     \end{array}
   \right.
\end{equation}
where $\bar{z}_{i} =(z_{i} -z_0)/d$, indicates the distance between the predictor redshift value for the $i$-th
observation and the focal redshift $z_0$. $d$ is the maximum distance between the point of interest
and elements inside the window. 

\textbf{Step A3}. \textit{Regression analysis.}  Following the Loess technique, 
we consider a low-degree polynomial to perform a local fit of the subsample in each window. It can be 
possible that higher-degree polynomials works, but for a simple nonparametric regression model we 
perform the analysis with a polynomial of the form
\begin{equation}\label{eq:polinomio1}
H(z)=a_0 +a_1 z.
\end{equation}
The first term correspond to the fitting coefficients of $H(z)$, which are
calculated by consider an evaluation in $z=0$ as $H(0)=a_0$. 
A similar fit routine proposal was presented in [\cite{Daly:2003iy}]. The r.h.s second term is
related to $H'$, parameter that we will reconstruct
in Sect. \ref{sec:loess-simex-model2}. The
reconstructed quantity is a weighted sum of the observations
$H(z)$ represented as: 
\begin{equation}
\hat{H}(z)=\sum\limits_{i=1}^{n} W_{ij}H_{i},
\end{equation}
where the weights in this regression are $W_{ij}=W[(x_{i}-x_{0})/d]$ and $j=1,\ldots,k$.

\textbf{Step A4}. \textit{Simulated data sample.} The Simex method offers an algorithm 
to estimate a \textit{true} parameter set in situations where covariate data has noise. Basically, this step 
consist in 
adding to the data sets an additional measurement 
error as follow
\begin{equation}\label{eq:simexH}
\eta_{i}(\lambda)=H_i +\sqrt{\beta}\sigma_{H_i},
\end{equation}
where $\eta_{i}(\lambda)$ denotes the simulated data points and $\sigma_{H_i}$ is the measurement 
error variance of each $H(z)$ observation. The resulting measurement error is 
$\beta=(1+\lambda)$, in where we can extrapolate the data sample to an \textit{error free zone}
if $\lambda =-1$. This zone is achieved after perform a standard regression, using a quadratic polynomial, 
of the data set computed for difference values of $\lambda$. Specifically, we are going to consider 
as a starting value $\lambda =0.5$ until $\lambda =2$ increasing in steps of $0.1$.

\textbf{Step A5}. \textit{Starting the reconstruction.} After performing the latter extrapolation step  
the data set will be simplified to the same length of the initial data and finally, these simulated data sets are 
normalized by $H_0$, given as a result the reconstruction of $h(z)$.
All the above steps are repeated for all the data points in the astrophysical sample.
The connection of the Loess-Simex reconstructed data points 
are represented by a curve due the lack of parameter estimates. The reconstructed normalized Hubble 
parameter $h(z)$ gives a general trend of the model.

\textbf{Step A6}. \textit{About the confidence regions}. To design the confidence 
regions of the reconstructed parameter $h(z)$ we require the transfer uncertainties via error propagation 
given by
\begin{eqnarray}
{\sigma_{h}}^2&=&\left(\frac{\sigma_{H}}{H_0}\right)^2 +\left(\frac{H^2}{{H_0}^4}\right)\sigma_{H_0}^2. \label{eq:errorsigma1}
\end{eqnarray}
With this expression we can calculate the uncertainties for the $O_{m}$ diagnostic
\begin{eqnarray}
\sigma^{2}_{{O^{(1)}_m}}&=&\left[\frac{2h}{z(3+3z+z^2)}\right]^2 {\sigma_{h}}^2. \label{eq:sigmaOm1}
\end{eqnarray}
For the dynamical $O_{m}$ diagnostic we have the following uncertainties
\begin{eqnarray}
\sigma^{2}_{{O^{(2)}_m}}&=& \left[\frac{-4h(1+z)+2z(2+z)h'}{z^2 (1+z)(3+z)}\right]^2 \sigma^{2}_{h} \nonumber\\&&
+\left[\frac{2(2+z)h}{z (1+z)(3+z)}\right]^2 \sigma^{2}_{h'}, \label{eq:sigmaOm2}
 \end{eqnarray}
 \begin{eqnarray}
\sigma^{2}_{O_k} &=& \left[\frac{6h(1+z)^2 -2z(3+3z+z^2)h'}{z^2 (1+z)(3+z)}\right]^2 \sigma^{2}_{h}\label{eq:sigmaOk} 
\nonumber\\&&
+\left[\frac{-2(3+3z+z^2)h}{z(1+z)(3+z)}\right]^2 \sigma^{2}_{h'}.
\end{eqnarray}

As the set implies, we need to found the value of the variable $\sigma_H$. Let us start with the fitted value
$\hat{H}(z)$ obtained in the \textbf{Step A3}.  For nonparametric regression
models we estimate the error variance as
\begin{equation}\label{eq:squareH}
S^2 =\frac{1}{n-df_{mod}}\sum\limits^{n}_{i=1}r^{2}_{i},
\end{equation}
where $r_{i}=H_{i}-\hat{H}_{i}$ is the residual for $i$-th observation and $df_{mod}$ is the
equivalent degrees of freedom for the model, which in our case it is equal to two. With this we are capable
to compute the variance of the fitted value $\hat{H}(z)$ at $z=z_{i}$ as:
\begin{equation}\label{eq:varianceH}
\hat{V}(\hat{H}_{i})\equiv\sigma^2_{\hat{H}_i} =S^2 \sum\limits^{n}_{j=1}W^2_{ij}.
\end{equation}
The results of the latter are considered to compute the propagation values $\sigma_h$ in  
Eq.(\ref{eq:errorsigma1}). Finally, the 68\% confidence interval and the 95\% confidence
interval are given by $h_{i} \pm \sqrt{\hat{V}(\hat{H}_{i})}$ and $h_{i} \pm 2\sqrt{\hat{V}(\hat{H}_{i})}$,
respectively and $h_i =\hat{H_i}/H_0$. 

\subsection{Reconstruction of $h'(z)$}
\label{sec:loess-simex-model2}
The logistics in this issue remains in the steps explained above. Nonetheless, we are 
going to proceed with a data set that only includes the coefficients related to the first derivative of 
$H(z)$. 

\textbf{Step B1}. \textit{Reconstruction of $h'(z)$}. Let us proceed as in \textbf{Step A1} until \textbf{Step A3}, 
where in the latter we performed a linear fit for these points using Eq.(\ref{eq:polinomio1}). 
The fitting coefficients of our interest are determinated by the evaluation of the polynomial in $z=0$ as 
$H'(0)=a_1$, where the prime denotes differentiation with respect to $z$. The new data set will consist of these 
$a_1$ coefficients for the 28-simulated data points, 
to which we apply a least square fit and then extrapolate to $\lambda =-1$, given us the data set
that we normalize to obtain the values of $h'(z)$ and its respectively curve as
in the \textbf{Step A5}.

\begin{figure*}
\begin{center}
\includegraphics[width=1.06\textwidth]{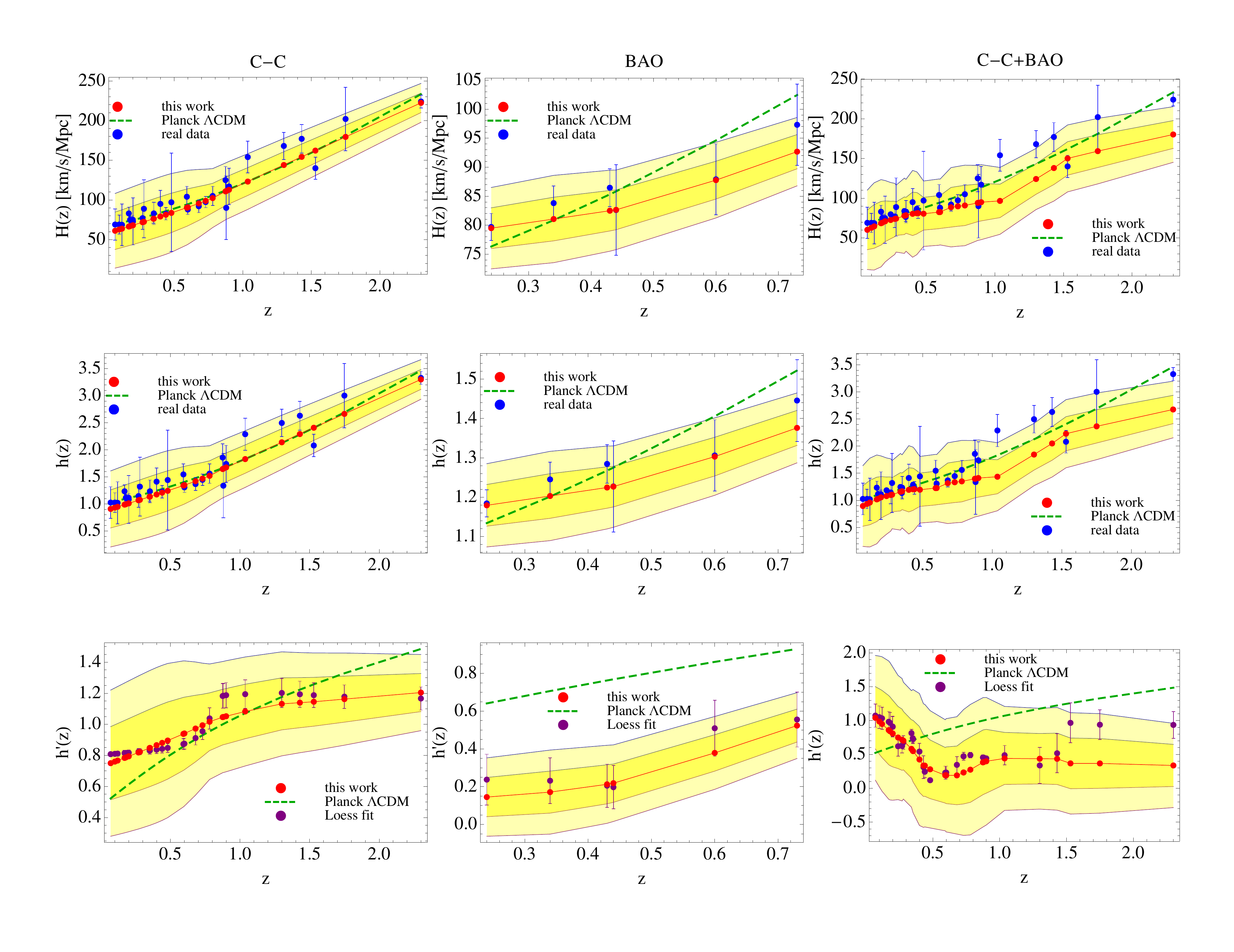}
\caption{Reconstruction of $H(z)$, $h(z)$ and $h'(z)$ parameters for C-C data (left column), BAO data (middle column) and C-C+BAO data (right column). The red dots(line) are(is) the Loess-Simex results for each sample. The dashed green line
is $\Lambda$CDM with $\Omega_m =0.315$. Shaded yellow areas represent the $68\%$ and $95\%$ confidence regions.
Top row: $H(z)$ Loess-Simex reconstructions. The blue dots are the real data sample with its
respectively error bars. Middle row: $h(z)$ Loess-Simex reconstructions. The blue dots represent the normalize real data $h$
with its respectively error propagation bars. Bottom row: $h'(z)$ Loess-Simex reconstructions. The purple dots represents the 
values of the second coefficient after perform Loess routine fit, which also gives the uncertainties bars via the covariance matrix.}
\label{fig:ploth-hp}
\end{center}
\end{figure*}

\textbf{Step B2}. \textit{About the error propagation}. Estimate the errors of $h'(z)$ 
and construct a similar step as was developed with 
Eqs.(\ref{eq:squareH})-(\ref{eq:varianceH}) can be a little tricky and it is necessarily to be careful
in the following methodology. 
This can be seeing from the form of Eq.(\ref{eq:simexH}),
expression that can be used similarly for $h'(z)$ if we have at hand the values of $H'(z)$ 
(already obtained in the linear fit performance in \textbf{Step B1}). The next 
question is:  how we can compute the uncertainties of $H'(z)$? We need to start from
\textbf{Step A4}, where we perform a least square fit and the polynomial that we need
to propagate now is
\begin{equation}
{\sigma_{H'}}^2=\sigma^2_{a_0} +z^2 \sigma^2_{a_1} +z^4 \sigma^2_{a_2},
\end{equation}
where the $\sigma$-values are the diagonal elements of the covariance matrix obtained
from $H'(z)$ data set.

With the new set $[H'(z),\sigma_{H'}]$, we are ready to reproduce the same steps starting 
in Eq.(\ref{eq:simexH}) and computing its error and matrix variance Eqs.(\ref{eq:squareH})-(\ref{eq:varianceH}). 
Until now we are not taken yet into
account any normalization of $H'(z)$, aspect that is implicit in the following propagation of errors
\begin{eqnarray}
{\sigma_{h'}}^2&=&\left(\frac{\sigma_{H'}}{H_0}\right)^2 +\left(\frac{H'^2}{{H_0}^4}\right)\sigma_{H_0}^2.\label{eq:errorsigma2}
\end{eqnarray}
Finally, using this error propagation and its respectively $h'(z)$ value we can construct
the \textit{confidence regions} as in \textbf{Step A6}.
\subsection{Nonparametric reconstruction of the $O_m$ diagnostic}

On one hand, regarding to the $O_m$ diagnostic for $\Lambda$CDM flat model (\ref{eq:omega1}), it is 
straightforward to compute the $O_m$ data set using the Loess-Simex estimates values 
$h(z)$ calculated in Sect. \ref{sec:loess-simex-model}. The values of $O_m$ are given directly 
from the new data set $\hat{h}(z)$.

On the other hand, the uncertainties calculations are 
easily to perform 
via Eq.(\ref{eq:sigmaOm1}). Thereupon, we
construct the 68\% and the 95\% confidence
intervals using the expressions: $\hat{O}_{m} \pm \sigma_{\hat{O}_m}$ and
 $\hat{O}_{m} \pm 2\sigma_{\hat{O}_m}$, respectively.

As we discussed, the existence of a non-flat universe brings to the scene
$h'(z)$ and $O_k$. 
In this case the system is given by Eqs.(\ref{eq:omega2})-(\ref{eq:omegak}),
which are independent of the values of the cosmological parameters $\Omega_m$ and $\Omega_k$
and implying a model that only relies in the values of our reconstructed $h(z)$ and $h'(z)$. 
 
The confidence regions will be compute using the error propagation Eqs.(\ref{eq:sigmaOm1})-(\ref{eq:sigmaOm2}) 
and the expressions: $\hat{O}^{(2)}_{m} \pm \sigma_{\hat{O}^{(2)}_m}$,
 $\hat{O}^{(2)}_{m} \pm 2\sigma_{\hat{O}^{(2)}_{m}}$ and $\hat{O}_{k} \pm \sigma_{\hat{O}_k}$, 
  $\hat{O}_{k} \pm 2\sigma_{\hat{O}_k}$.

\begin{figure*}
\begin{center}
\includegraphics[width=1.06\textwidth]{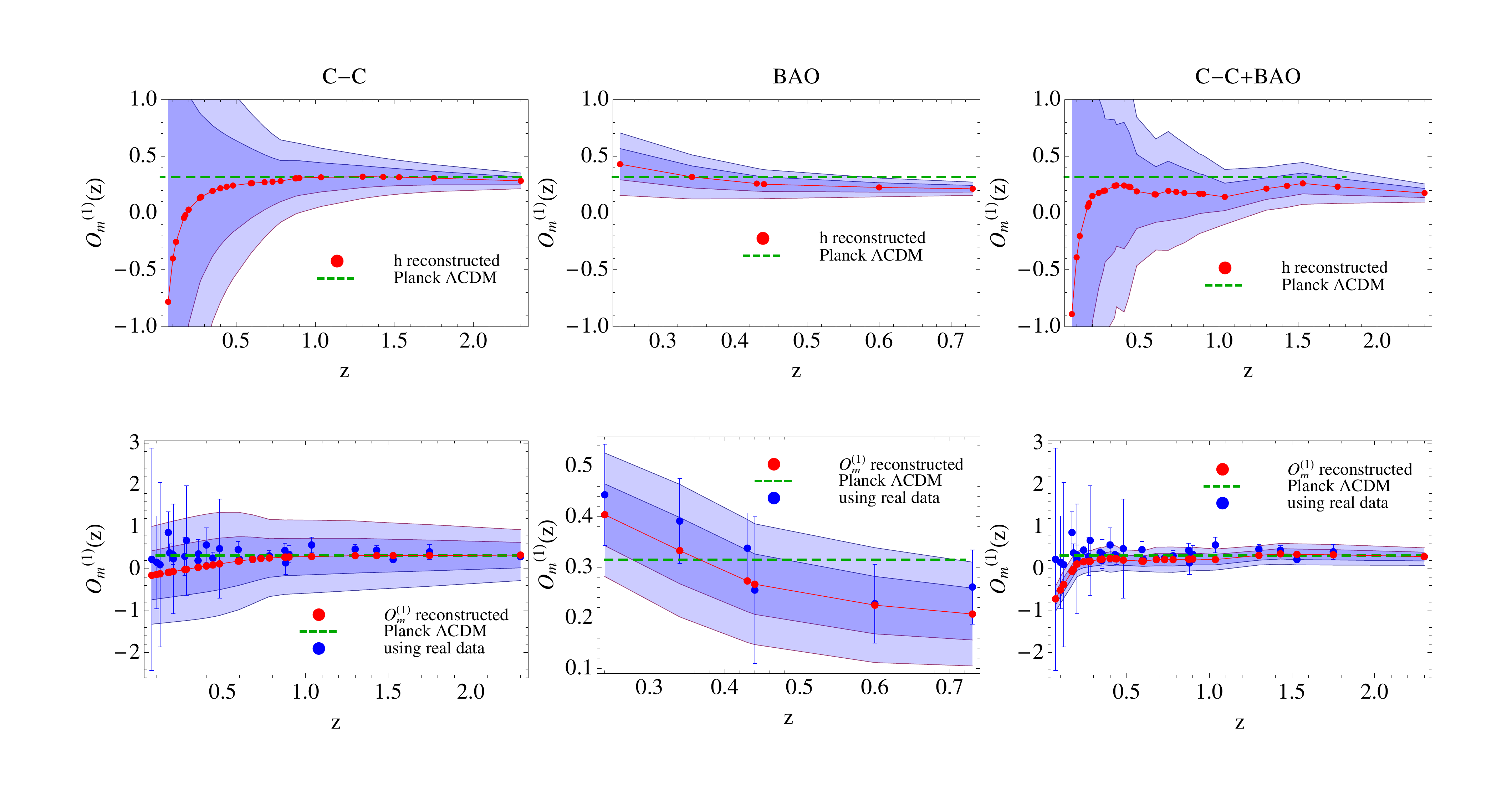}
\caption{Reconstruction of the $O^{(1)}_{m}$ diagnostic for C-C data (left column), BAO data (middle column) and C-C+BAO data (right column).
The red dots(line) are(is) the Loess-Simex results for each sample. The dashed green line
is $\Omega_m =0.315$. Shaded purple areas represent the $68\%$ and $95\%$ confidence regions.
Top row: $O^{(1)}_{m}$ diagnostic with $h$ reconstructed via Loess-Simex. Bottom row: $O^{(1)}_{m}$ values reconstructed 
\textit{directly} via Loess-Simex. The
blue dots are these values using $h$ normalized with its error propagation bars.}
\label{fig:ploth-OMrecons}
\end{center}
\end{figure*}

\begin{figure*}
\begin{center}
\includegraphics[width=1.06\textwidth,origin=c,angle=0]{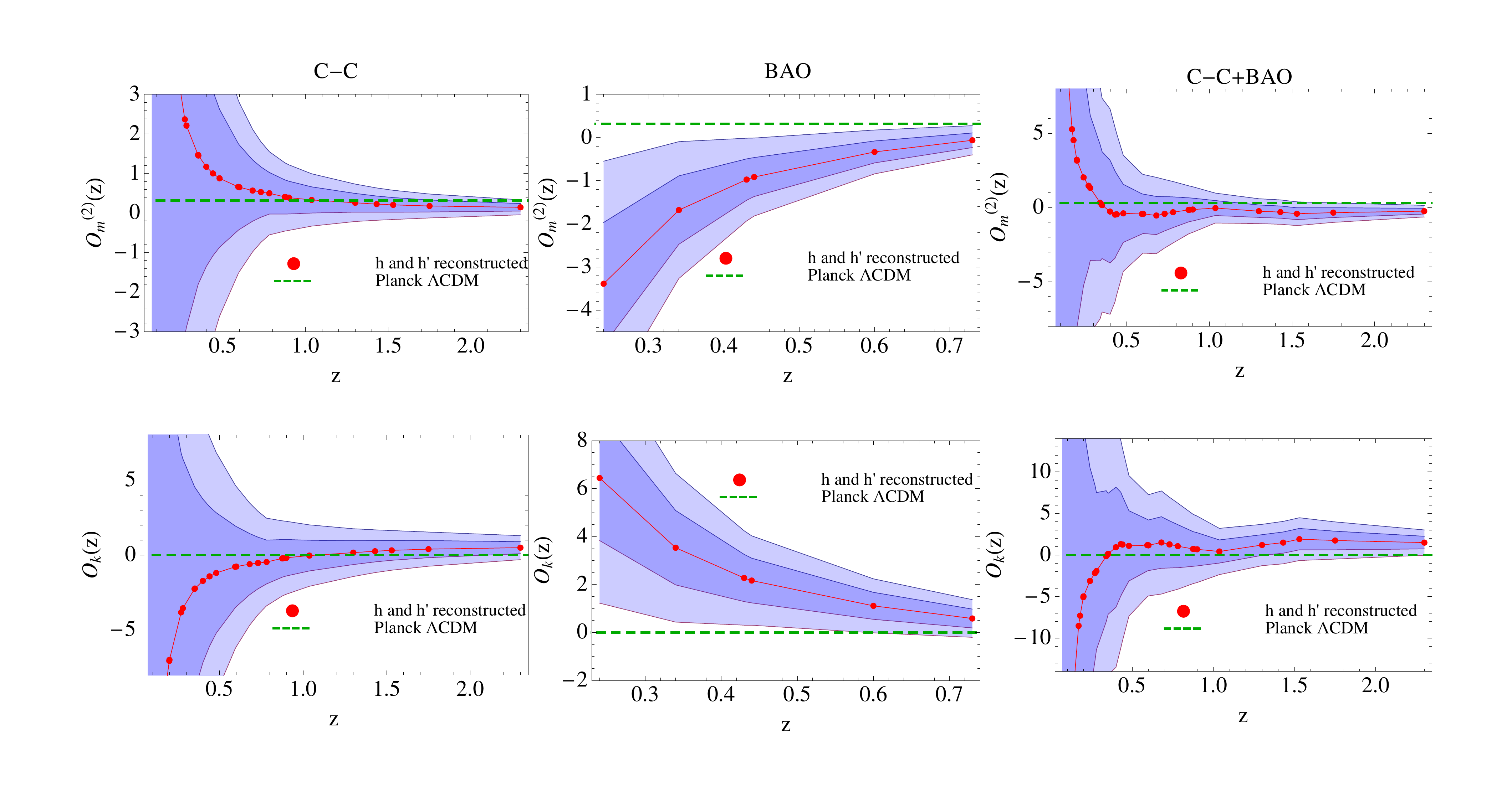}
\caption{Reconstruction of the $O^{(2)}_{m}$ and $O_k$ diagnostics for C-C data (left column), BAO data (middle column) and C-C+BAO data (right column). The red dots(line) are(is) the reconstructed $O^{(2)}_{m}$ values using the reconstructed $h$ and $h'$ via Loess-Simex. The dashed green line is $\Omega_m =0.315$. Shaded purple areas represent the $68\%$ and $95\%$ confidence regions.
Top row: $O^{(2)}_{m}$ diagnostic with $h$ and $h'$ reconstructed via Loess-Simex. Bottom row: $O_{k}$ diagnostic with $h$ and $h'$ reconstructed via Loess-Simex.}
\label{fig:ploth-OMrecons2}
\end{center}
\end{figure*}

\section{Discussion and Conclusions}
\label{sec:interpretation}

We developed the Loess-Simex factory to achieve two interesting goals. First, we perform the reconstruction  
of the normalized Hubble parameter $h(z)$, results that are represented by 
red dots (red line) in Figures \ref{fig:ploth-hp}, \ref{fig:ploth-OMrecons} 
and \ref{fig:ploth-OMrecons2}. As well, in the upper plots
of Figure \ref{fig:ploth-hp} we illustrate the original $H(z)$ data set represented by blue dots with its respectively 
error values and its nonparametric reconstruction (red dots/line). It is interesting to remark the comparison between 
these reconstructed points and the $\Lambda$CDM model,
which is represented by a dotted green line.

Our second goal was the reconstruction of the $O_m$ diagnostic 
and the $O^{(2)}_m$ and $O_k$ parameters using two astrophysical samples (C-C and BAO) for $H(z)$ and the combination of them. 
The reconstruction of 
$O_m$ diagnostic was made by consider two options: \textbf{(I)} using the already reconstructed $h$ 
values (top of Figure \ref{fig:ploth-OMrecons}) and \textbf{(II)} performing \textit{directly} its reconstruction 
(bottom of Figure \ref{fig:ploth-OMrecons}).

\begin{figure*}
\begin{center}
\includegraphics[width=7.6cm]{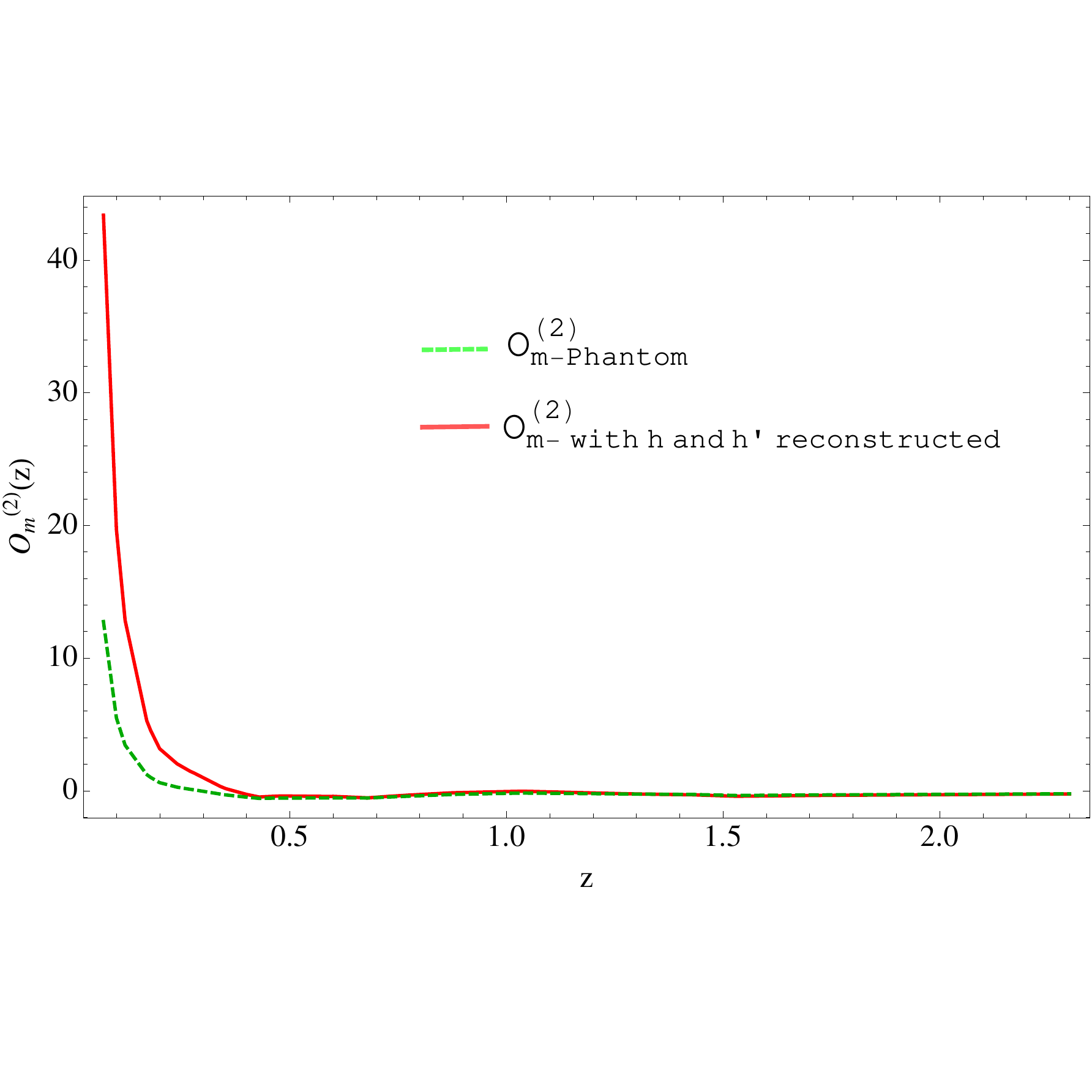}
\includegraphics[width=7.6cm]{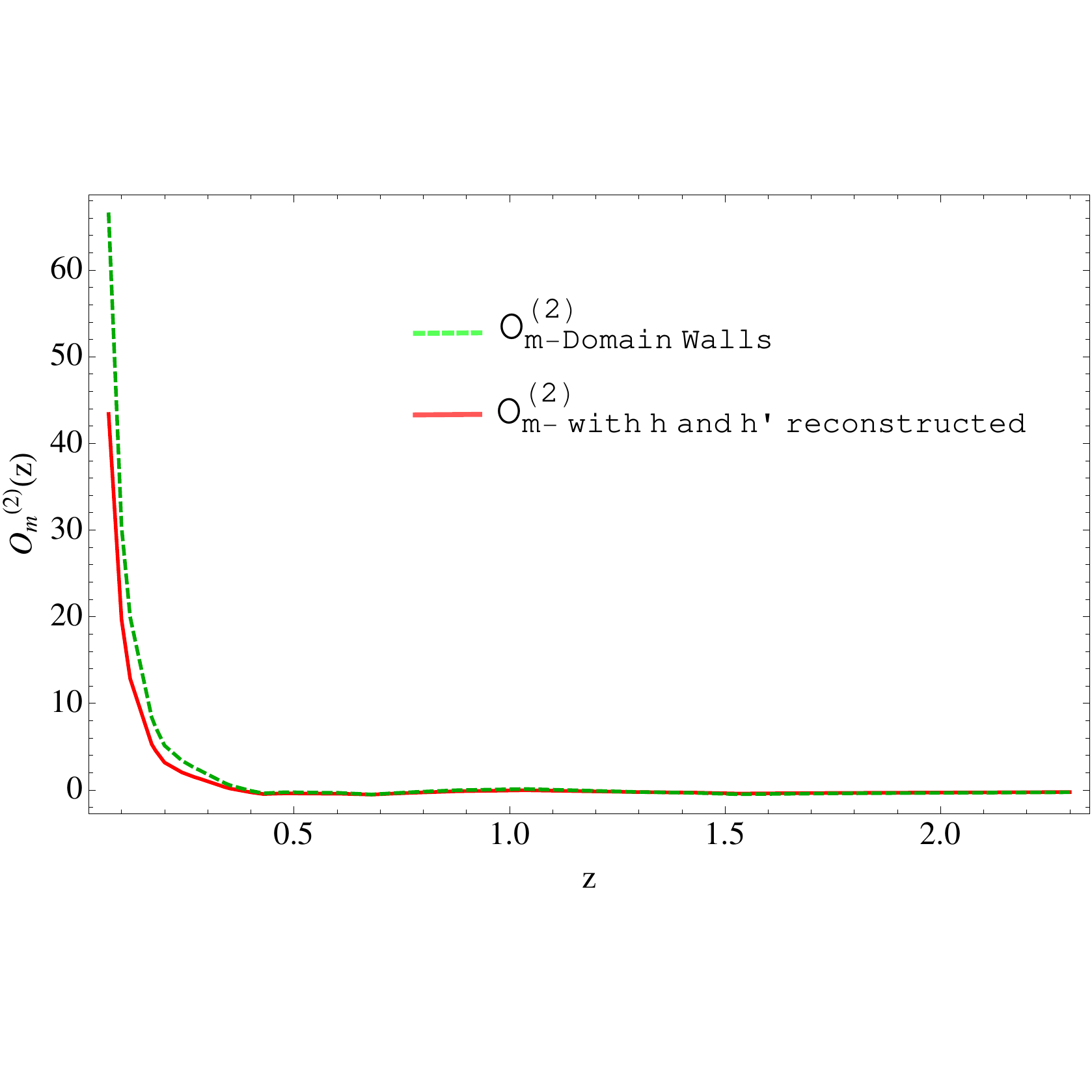}
\includegraphics[width=7.6cm]{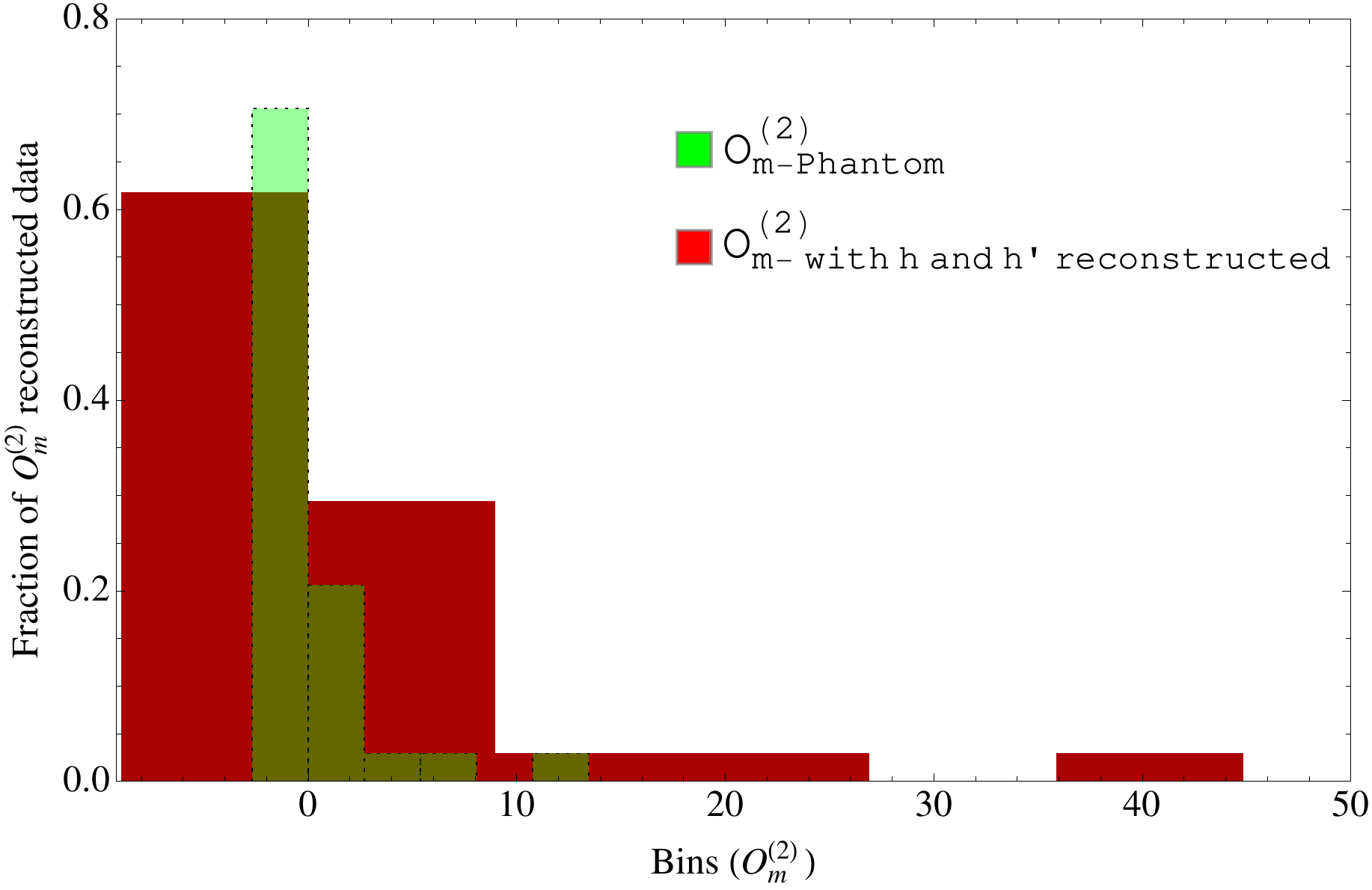}
\includegraphics[width=7.6cm]{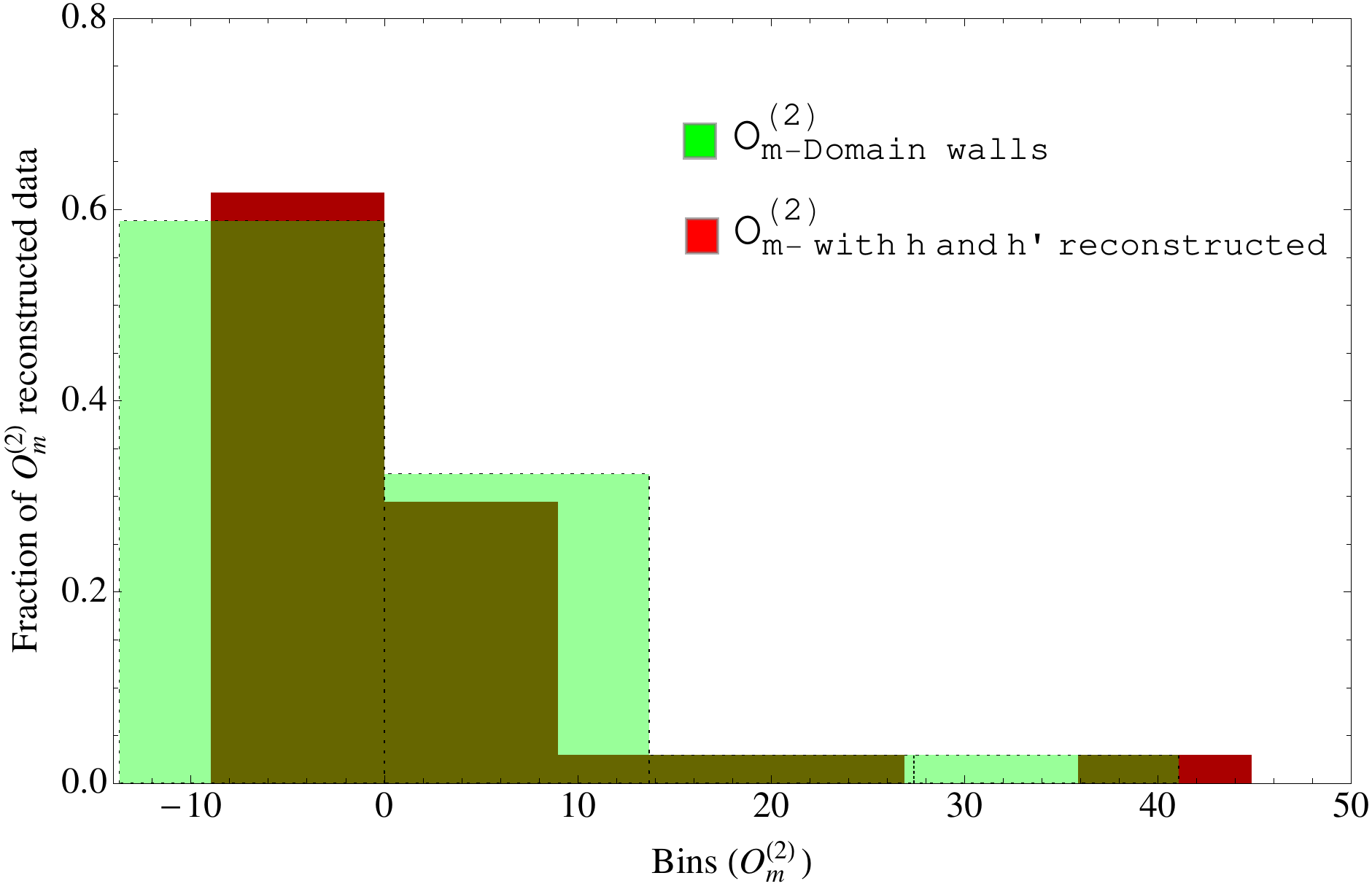}
\caption{\textit{Top:} Comparison between DE models: phantom and static domain walls and the $O^{(2)}_{m}$ reconstructed using C-C+BAO data.  
The green dashed line represent Eq.(\ref{eq:omgen2}) for a phantom ($w_{0} <-1$) and static domain walls ($w_{0} =-2/3$) models . The red solid line represent the  $O^{(2)}_{m}$ diagnostic using the reconstructed $h$ and $h'$. We observe at the right hand side plot that the static domain walls model appears to be more in agreement with the $O^{(2)}_{m}$ diagnostic reconstructed in comparison to the left plot where the phantom model starts to deviate from the $O^{(2)}_{m}$ reconstructed at low redshifts ($z<0.5$).
\textit{Bottom:} Probability comparison between DE models. The green bars represent DE models (phantom and static domain walls) and the red bars represent the amount of the reconstructed data. $62\%$ fraction of the reconstructed data lies in $O^{(2)}_{m}<0$, then in this range we observe that the amount of $O^{(2)}_{m}$ deviation between this data and each DE models correspond to $8\%$ for a phantom model and $3\%$ for a static domain walls. These probabilities supports the result obtained above.}
\label{fig:ploth-modelscomparison}
\end{center}
\end{figure*}

Let us discuss the results for each case.

For the C-C sample, the nonparametric reconstruction has the same trend as
the one reported in [\cite{Montiel2013}]. However, in our case we worked with the normalized 
Hubble parameter $h$, which behaviour is analogous to the previous case, as it is expected.
The \textit{directly} reconstruction of the $O_m$ diagnostic appears to be in good agreement with $\Lambda$CDM at $z>1$.
It is interesting to notice that in this case 
the confidence regions looks smaller than in the case when we use the $h$ data reconstructed.

For the BAO sample, unlike other proposals above mentioned, ours results shows a $\Lambda$CDM model 
that lies in our $O_m$ confidence contour reconstructions at 2-$\sigma$, even by performing the reconstruction 
with a few values of this data set.  As in the
previous sample, the \textit{directly} reconstruction of this diagnostic gave a 
concordance model between $1$ up to 2-$\sigma$. 
The reconstructions
of $O^{(2)}_{m}$ and $O_k$ implies the reconstruction of $h'$ and the analysis shows large uncertainties,
even so, the reconstructions at high redshifts shows a trend that possibly can loiters to $\Lambda$CDM at $z>0.7$ 
(see Figure \ref{fig:ploth-OMrecons2}, middle row).

For the C-C+BAO sample, we observe that the reconstruction is almost similar to the C-C case,
clearly due the amount of data of the first sample in comparison to the second sample. 
The concentration of data points at $z<0.5$ is related to the effects of the evaluation of the reconstructed data in
Eq.(\ref{eq:omega1}). We observed in the $O^{(2)}_{m}$ analysis a pull of the reconstructed curve up at $z<0.3$, which probably
shows the important relationship between derivatives of the data and the model itself.
The \textit{directly} reconstruction at zero-order loiters to $\Lambda$CDM up to $z=1$, but due that is not a constant 
in the entire redshift range we need to consider a dynamical test. 

In order to found the adequate DE model in agreement with the reconstructions we
perform a $O^{(2)}_{m}$ diagnostic (first-order in $h$, i.e $h'$) finding that even when the $O^{(1)}_{m}$ diagnostic 
hint a phantom
behaviour, when we enter in the region $w> -1$ the reconstructions have a preference for a EoS with known physical meaning $w=-2/3$, which correspond to a static domain wall network in the entire redshift range.
At the top of Figure \ref{fig:ploth-modelscomparison} we compare the 
dynamical $O^{(2)}_{m}$ diagnostics reconstructed (red curves) with  $O^{(2)}_m$ diagnostics Eq.(\ref{eq:omgen2})
(green dashed curves) using two specific DE EoS models.
How much is the fraction of the reconstructed data that make one DE model better than the other?
To answer this, we calculated the probability of this fraction for each DE model in terms of the $O^{(2)}_{m}$ bins. The results are
represented by the histograms at the bottom of Figure \ref{fig:ploth-modelscomparison}. The green bars represent DE models (phantom and static domain walls) and the
red bars represent the amount of the reconstructed data. The bin width for the $O^{(2)}_{m}$ reconstructed values are calculated by using [\cite{scott-rule}]. We have that $62\%$ of the reconstructed data lies in $O^{(2)}_{m}<0$, then in this range we observe that the amount of $O^{(2)}_{m}$ deviation between this data and each DE models correspond to $8\%$ for a phantom model and $3\%$ for a static domain walls, making the latter a better model in agreement with the reconstructed data.

Forthcoming studies along the lines of these analysis promise to greatly improve with the use of high quality observations to
make this nonparametric $O_m$ diagnostic more accurate and a very useful tool for testing alternatives
DE parameterizations and modify gravity proposals.

{\bf Acknowledgements:} C. E-R thank CNPq for financial support and would like to thank R. Lazkoz and J. Alcaniz for their opinions along these ideas. J.C.F.  would like to thank FAPES and CNPq (Brazil) for financial support.

\appendix
\section{Reconstruction of $D(z)$}\label{GE}

In order to formulate a test for DE models, let us consider the derivative of the luminosity distance 
(\ref{eq:lum_dist1}) and the distance-redshift (\ref{eq:dist-red1})
to obtain the following expressions:
\begin{eqnarray}
d'_L &=& \frac{c}{H_0\sqrt{-\Omega_k}}\sin{\left(\sqrt{-\Omega_k}\int^{z}_{0}dz' \frac{H_0}{H(z')}\right)} 
\nonumber\\&&
+\frac{c(1+z)}{H}\cos{\left(\sqrt{-\Omega_k}\int^{z}_{0}dz' \frac{H_0}{H(z')}\right)},\label{eq:ddL}
\end{eqnarray}
\begin{eqnarray}
D'(z)&=& \frac{H_0}{c}\left[-\frac{d_L}{(1+z)^2}+\frac{d'_L}{(1+z)}\right].\label{eq:DD}
\end{eqnarray}
From where we can extract the following cases:
\begin{itemize}
\item If we have a flat universe ($\Omega_k =0$) then the equations are
\begin{eqnarray}
d'_L&=&\frac{c(1+z)}{H},\quad
D'=\frac{H_0}{H}\equiv h^{-1}.
\end{eqnarray}
\item For the case of a non-flat universe ($\Omega_k \neq 0$) we have
\begin{eqnarray}
d'_L &=& \frac{c}{H_0\sqrt{-\Omega_k}}\sin{\left(\sqrt{-\Omega_k}\int^{z}_{0}dz' \frac{H_0}{H(z')}\right)}
\nonumber\\&&
+\frac{c(1+z)}{H}\cos{\left(\sqrt{-\Omega_k}\int^{z}_{0}dz' \frac{H_0}{H(z')}\right)}, \nonumber \\
D'&=&\frac{H_0}{H}\cos{\left(\sqrt{-\Omega_k}\int^{z}_{0}dz' \frac{H_0}{H(z')}\right)}.
\end{eqnarray}
\end{itemize}
From Eq.(\ref{eq:friedmann}) we can obtain an expression for the derivative of the distance-redshift
\begin{equation}
D'^{-2}=\Omega_m (1+z)^3 +\Omega_k (1+z)^2 +(1-\Omega_m -\Omega_k)f(z),
\end{equation}
where $f(z)$ is given by (\ref{eq:fz}) which is $f(z)=1$ if $w_0 =-1$ and 
$f(z)=(1+z)^{3(1+w_0)}$ for a constant EoS.
Possible scenarios are:
\begin{itemize}
\item For $w=-1$ and $\Omega_k =0$,
\begin{equation}
D'^{-2}=\Omega_m (1+z)^3 +(1-\Omega_k).
\end{equation}
\item For $w=w_0$ and $\Omega_k =0$,
\begin{equation}\label{eq:gen}
D'^{-2}=\Omega_m (1+z)^3 +(1-\Omega_m)(1+z)^{3(1+w_0)}.
\end{equation}
\item For $w=-1$ and $\Omega_k \neq0$,
\begin{equation}
D'^{-2}=\Omega_m (1+z)^3 +\Omega_k (1+z)^2 +(1-\Omega_m -\Omega_k).
\end{equation}
\item For $w=w_0$ and $\Omega_k \neq0$,
\begin{eqnarray}
D'^{-2}&=&\Omega_m (1+z)^3 +\Omega_k (1+z)^2 \nonumber\\&&
+(1-\Omega_m -\Omega_k)(1+z)^{3(1+w_0)}. 
\end{eqnarray}
\end{itemize}
From Eq.(\ref{eq:gen}) we obtain the first generalized equation for the $O_m$ diagnostic described
by Eq.(\ref{eq:omgen}).

When we consider a non-flat universe the $\Omega_k$ arise and we are going to need a system of two
equation: the first one given by Eq.(\ref{eq:hdef2}) and the second is the EoS when we rearranged
Eq.(\ref{eq:de_eos}).
After straightforward calculations and redefining $\Omega_m \equiv O^{(2)}_{m}$ and $\Omega_k \equiv O_k$, we obtain
the generalized equations for a non-flat universe and a constant dark energy EoS described by Eqs.(\ref{eq:omgen2})-(\ref{eq:omgen2_1}).


\label{lastpage}
\end{document}